\documentclass[conference]{IEEEtran}
\IEEEoverridecommandlockouts
\usepackage{cite}
\usepackage{amsmath,amssymb,amsfonts}
\usepackage[linesnumbered, ruled]{algorithm2e}
\usepackage{graphicx}
\usepackage[binary-units]{siunitx}
\usepackage{subcaption}
\usepackage[inline]{enumitem}
\usepackage{tikz}
\usepackage{pgfplots,filecontents}
\usepackage{scalefnt}
\pgfplotsset{compat=newest}
\usepackage{filecontents}
\usepgfplotslibrary{units}
\usepackage{xcolor}
\usepackage{wrapfig}
\def\BibTeX{{\rm B\kern-.05em{\sc i\kern-.025em b}\kern-.08em
    T\kern-.1667em\lower.7ex\hbox{E}\kern-.125emX}}
\pgfplotsset{%
   every tick label/.append style = {font=\small},
   every axis label/.append style = {font=\small},
   every mark/.append style={scale=100},
   width=6.1cm,compat=1.3
}
\graphicspath{{figures/}{figures/examples/}{figures/results/}}
\definecolor{s1}{RGB}{228, 26, 28}
\definecolor{s2}{RGB}{55, 126, 184}
\definecolor{s3}{RGB}{77, 175, 74}
\definecolor{s4}{RGB}{152, 78, 163}
\definecolor{s5}{RGB}{255, 127, 0}
\definecolor{s6}{RGB}{25, 127, 0}
\definecolor{s7}{RGB}{25, 127, 55}
\definecolor{s8}{RGB}{255, 127, 45}
\definecolor{s9}{RGB}{85, 127, 65}
\definecolor{s10}{RGB}{0, 0, 255}
\definecolor{s11}{RGB}{0, 0, 0}
\definecolor{s12}{RGB}{255, 0, 0}

\pgfplotscreateplotcyclelist{set2}{
    s1, style={fill=s1}\\
    s2, style={fill=s2}\\
    s8, style={fill=s8}\\
    s9, style={fill=s9}\\
}

\pgfplotscreateplotcyclelist{set24}{
    s1,every mark/.append style={fill=s1},mark=*,mark size=1.5pt\\
    s2,every mark/.append style={fill=s2},mark=o,mark size=1.5pt\\
    s3,every mark/.append style={fill=s3},mark=triangle*,mark size=1.5pt\\
    s4,every mark/.append style={fill=s4},mark=square,mark size=1.5pt\\
    s5,every mark/.append style={fill=s5},mark=diamond*,mark size=1.0pt\\
}

\pgfplotscreateplotcyclelist{set28}{
    s1,every mark/.append style={fill=s1},mark=*,mark size=2.5pt, only marks\\
    s2,every mark/.append style={fill=s2},mark=o,mark size=2.5pt, only marks\\
    s3,every mark/.append style={fill=s3},mark=triangle*,mark size=2.5pt, only marks\\
    s4,every mark/.append style={fill=s4},mark=square,mark size=2.5pt, only marks\\
    s5,every mark/.append style={fill=s5},mark=diamond*,mark size=2.5pt, only marks\\
}

\begin{document}

\title{Multilayer Resource-aware Partitioning for Fog Application Placement
}

\author{\IEEEauthorblockN{Zahra Najafabadi Samani, Nishant Saurabh, Radu Prodan}
\IEEEauthorblockA{\textit{Institute of Information Technology, University of Klagenfurt, Austria}
\\zahra,nishant,radu@itec.aau.at
}}

\maketitle

\begin{abstract}
Fog computing emerged as a crucial platform for the deployment of IoT applications. The complexity of such applications require methods that handle the resource diversity and network structure of Fog devices, while maximizing the service placement and reducing the resource wastage. Prior studies in this domain primarily focused on optimizing application-specific requirements and fail to address the network topology combined with the different types of resources encountered in Fog devices. To overcome these problems, we propose a multilayer resource-aware partitioning method to minimize the resource wastage and maximize the service placement and deadline satisfaction rates in a Fog infrastructure with high multi-user application placement requests. Our method represents the heterogeneous Fog resources as a multilayered network graph and partitions them based on network topology and resource features. Afterwards, it identifies the appropriate device partitions for placing an application according to its requirements, which need to overlap in the same network topology partition. Simulation results show that our multilayer resource-aware partitioning method is able to place twice as many services, satisfy deadlines for three times as many application requests, and reduce the resource wastage by up to $15-32$ times compared to two availability-aware and resource-aware state-of-the-art methods.
\end{abstract}


\textcolor{red}{2021  IEEE.  Personal  use  of  this  material  is  permitted.  Permission  from  IEEE  must be  obtained  for  all  other  uses,  in  any  current  or  future  media,  including  reprinting/republishing  this  material  for  advertising  or  promotional  purposes,  creating  new collective works, for resale or redistribution to servers or lists, or reuse of any copyrighted component of this work in other works.}

\section{Introduction}
\label{sec:intro}
The Cloud-assisted Internet of Things (IoT) emerged as an essential accelerator of the fourth industrial revolution~\cite{taneja2017resource}. However, the rapid growth of IoT applications with multiple services makes it challenging for Clouds to satisfy their low-latency real-time requirements~\cite{hong2019resource} and deadline constraints. To mitigate these challenges, \emph{Fog computing}~\cite{ieee2018ieee} emerged as a crucial platform consisting of a large number of heterogeneous and geographically distributed resources, hierarchically split between the Cloud and the IoT resources. This enables the Fog to address latency-centric constraints posed by the Cloud through systematic placement~\cite{oueis2015small} of services across resources closer to the IoT devices. However, the non-uniform Fog network composed of heterogeneous devices (e.g., routers, switches, gateways) significantly varies in terms of processing speed, network bandwidth and storage capacity~\cite{hong2019resource}, which induces service availability and deadline fulfillment challenges for time-critical IoT applications.

To solve IoT application placement issues in Fog, there is a need to research strategies that increase its resource utilization, while fulfilling diverse application requirements. Prior researches leveraged evolutionary multi-objective optimization algorithms~\cite{sun2018multi}, and linear programming~\cite{velasquez2017service,huang2014co} models to identify cost-efficient IoT application placement strategies. Such approaches primarily focused on optimizing energy consumption, network usage and response time of IoT service requests, but limited to application specific requirements without considering resource dependencies among interrelated services. Other works~\cite{sajjad2015smart,shooshtarian2019clustering,lera2018availability,filiposka2018community, asensio2020designing} improved upon these studies and explored resource partitioning at a network topology level without considering other resources of Fog devices, such as processing speed, memory or storage sizes. Moreover, these methods failed to address the resource wastage, while optimizing the utilization of capacity-constrained Fog resources.

We approach this problem using a \emph{multilayer resource-aware partitioning} method that handles the resource diversity and the interconnection structure of Fog devices to minimize the resource wastage and optimize the application service placement. To address the resource diversity, we model the heterogeneous Fog infrastructure as a multilayer graph comprising the network topology and heterogeneous devices with different CPU speed, memory and storage capacities. To optimize the application placement, we split the Fog devices in overlapping partitions with respect to the network topology and different resource types. This enables our method to target the correct set of resources for an application and improve its availability. Afterwards, the placement involves two steps.

\paragraph{Feature partition selection} matches the requested application services with partitions based on their resource requirements, which need to overlap in the same network topological partition underneath.

\paragraph{Service placement} maps services onto the appropriate Fog devices in the elected partitions, such that all application services reside in the same network partition.

Extensive simulation experiments demonstrate that our multilayer resource-aware partitioning method is able to place up to twice as many services, satisfy deadlines for up to three times as many application requests, and reduce the Fog resource wastage by up to $15-32$ times compared to two related state-of-the-art methods~\cite{lera2018availability,taneja2017resource}.

The paper is organized in seven sections. Section~\ref{sec:sota} summarizes the related work. Section~\ref{sec:model} presents the model underneath our method, including the multilayered Fog representation and its key layer partitioning method in section~\ref{sec:partition}. Section~\ref{sec:placement} describes multilayer Fog placement method, including two feature partition selection and service topology optimization algorithms. Section~\ref{sec:eval} presents the experimental results and Section~\ref{sec:conclusion} concludes the paper.

\section{Related work}
\label{sec:sota}
This section revisits the recent related works on Fog application placement across the following categories.   
\paragraph{Application-aware placement} Huang et al.~\cite{huang2014co} proposed an integer programming model to optimize the energy cost by merging neighboring services onto a single device in a multi-hop Fog network. Oueis et al.~\cite{oueis2015small} formulated a resource allocation model in Fog to jointly optimize power consumption and latency by clustering resources and assigning each cluster to a requested application service. Velasquez et al.~\cite{velasquez2017service} designed an integer linear programming placement for IoT services across a Cloud-Edge infrastructure that optimizes service latency.
Naha et al.~\cite{naha2020deadline} proposed a resource provisioning algorithm for deadline-based application placement in Fog to optimize processing time, processing cost and network delay. These approaches primarily focused on optimizing application-specific requirements without considering the network topology and resource heterogeneity.

\paragraph{Topology-aware placement} Filiposka et al.~\cite{filiposka2018community} designed a community-based Fog management that exploits distributed hierarchical clustering to reduce latency and optimize service migration. Similarly, Asensio
et al.~\cite{asensio2020designing} proposed a distributed control and management approach that groups Fog resources based on their network topology to minimize latency and energy consumption.
In contrast, Sun et al.~\cite{sun2018multi} modeled the resource placement in Fog as a multi-objective execution and latency optimization that computes different service clusters for an application. 
These approaches utilized clustered Fog resources based on network topology without considering the infrastructure heterogeneity.

\paragraph{Resource-aware placement} Shooshtarian et al.~\cite{shooshtarian2019clustering} proposed a two-phase allocation method that hierarchically represents Fog resources and performs local clustering in each layer to optimize resource utilization and network delay. Taneja et al.~\cite{taneja2017resource} proposed a resource-aware application mapping approach to optimize resource utilization in Fog. Similarly, Stefanic et al.~\cite{stefanic2020quality} proposed a subgraph pattern matching approach for application placement that maps the multi-tier application graph onto the Fog infrastructure to improve resource utilization. Nevertheless, these three approaches only consider different Fog resources characteristics and ignore the network topology structure of Fog devices. Contrarily, Lera et al.~\cite{lera2018availability} proposed a greedy approach for application placement that optimizes availability and latency by partitioning Fog resources in hierarchical clusters based on their connectivity. However, it fails to minimize resource wastage by ignoring Fog resource characteristics during the greedy-based assignment.
\section{Model}
 \label{sec:model}
This section presents a formal model essential to this work.
\subsection{Resource infrastructure}
\label{subsec:resource}
A \emph{Fog infrastructure} consists of three layers:

\subsubsection{Cloud layer} represents a data center with high-performance computing resources.

\subsubsection{Fog network layer} $F = \left(D, N\right)$ lies between the Cloud and the end-users, and provides close proximity computational and storage services on top of two resource sets, according to the architecture proposed by the OpenFog consortium~\cite{ieee2018ieee}:

\paragraph{Physical devices} $D= \left\{d_1, d_2, \ldots, d_n\right\}$, modeled as a triplet of resources $d_i = \left(R_{i1}, R_{i2}, R_{i3}\right)$, where $R_{i1}$ represents the speed of a CPU core in millions of instructions (MI) per second, $R_{i2}$ represents the memory size in \SI{}{\giga\byte}, and $R_{i3}$ represents the storage size of a device $d_i$ in \SI{}{\tera\byte}.

\paragraph{Network connections} $N = \left\{ n_{ij} | \left(d_i, d_j\right) \in D \times D\right\}$ between a subset of devices, where a network connection $n_{ij}= \left(BW_{ij}, LAT_{ij}\right)$ depends on the bandwidth $BW_{ij}$ and the latency $LAT_{ij}$ between the devices $d_i$ and $d_j$. 

\subsubsection{Client layer}
consists of a set of users $\mathcal{U} = \left\{u_1, \ldots, u_q\right\}$, including sensor and actuator client devices that request Fog resources for placing their applications. We do not consider user mobility in this work.

\subsection{Application model}
\label{subsec:appmodel}
We model an \emph{application} $A = \left(\mathcal{S}, M, \theta, u\right)$ requested by a user $u$ as a directed graph of \emph{services} $\mathcal{S} = \left\{s_1, s_2, \ldots, s_m\right\}$ interconnected through request \emph{messages} $M$. Every service $s_i \in \mathcal{S}$ has a triplet of resource demands $s_i = \left(r_{i1}, r_{i2}, r_{i3}\right)$, where $r_{i1}$ is the workload in MI, $r_{i2}$ is the required memory size, and $r_{i3}$ is the storage size.

We model each \emph{request message} $m_{ij} \in M$ based on its size $SZ_{ij}$, its source $s_{i} \in \mathcal{S}$, and its destination service $s_{j} \in \mathcal{S}$: $m_{ij} = \left(SZ_{ij},s_{i},s_{j}\right)$. A user $u\in\mathcal{U}$ triggers the application execution via an \emph{initial request message} $m_{ui}$ to service $s_i$. 

The application also has a completion \emph{deadline} that requires the completion of all its services.

\subsection{Multilayer Fog model}
\label{subsec:fog-model}
The Fog network layer represents the topological interconnection of the physical devices and does not capture their heterogeneous resources. To better handle resource diversity for application placement, we model the Fog across four layers $L=\left\{l_0, l_1, l_2, l_3\right\}$ representing relationships among its devices based on the network topology and three resource types.
\begin{enumerate}[font=\emph,wide]
    \item[(i)]\emph{Network layer $l_0$} corresponds to the Fog network layer modelled in Section~\ref{subsec:resource}.
    \item[(ii)--(iv)]\emph{CPU layer $l_1$, memory layer $l_2$, and storage layer $l_3$} indicate similarities among the Fog devices according to their CPU speed, memory size, and storage size.
\end{enumerate}

We model the Fog as a \emph{fully interconnected multilayer graph} $G = \left(D, \mathcal{E}, L\right)$, where $D$ is the set of Fog devices replicated across all four layers L, and $\mathcal{E} = \left\{ E_{ll^\prime}\ |\ \forall l, l^\prime \in [0, L)\right\}$
is the set of weighted bidirectional graph edges of two types.

\paragraph{Inter-layer edges} $E_{ll^\prime}=\left\{\left(d_i,d_i\right)\in D \times D\right\}$ connect each device $d_i$ in the layer $l \in L$ with the corresponding device $d_i$ in all the other layers $l^\prime \in L, l \neq l^\prime$. They indicate the connection between different resource characteristics of the same device and uncover the relation among resource types;

\paragraph{Intra-layer edges} $E_{ll} = \left\{\left(d_i, d_j\right) \in D \times D | i\neq j \right\}$ connect two Fog devices inside one layer $l \in L$ using a weight function $w^{(l)}: E_{ll} \to \mathbb{R}$, representing their \emph{similarity score}: 
\[w^{(l)}_{ij} = \frac{1}{{1}+{d_l\left(d_i,d_j\right)}},\]
where $d_l\left(d_i,d_j\right)$ is the Euclidean distance between their resource characteristics in layer $l \geq 1$:
$d_l\left(d_i,d_j\right) = R_{il} - R_{jl}$.
The Fog devices with a similarity score of 1 are exactly similar with respect to a resource $R_l$.

\subsection{Problem statement}
\label{subsec:probstate}
We introduce a number of definitions introducing our problem statement of placing an application $A = \left(\mathcal{S}, M, \theta, u\right)$ in a Fog environment $F=(D,N)$.

\subsubsection{Application placement} is a function $\mu: \mathcal{S} \to D \cup \varnothing$, where $\mu\left(s_i\right) = d_j$ satisfies the constraints for placing each service $s_i= \left(r_{i1},r_{i2},r_{i3}\right)$ on a Fog device $d_j=\left(R_{j1},R_{j2},R_{j3}\right)$: $\frac{r_{i1}}{R_{j1}}\le\theta$, $r_{i2} \leq R_{j2}$ and $r_{i3} \leq R_{j3}$. An \emph{invalid placement} $\mu\left(s_i\right)=\varnothing$ indicates that there exist no device in $D$ that satisfies the service constraints.

\subsubsection{Execution time} of a service $s_i$ is the ratio between its workload $r_{i1}$ and the speed $R_{i1}$ of the underlying hosting device $d_j$=$\mu\left(s_i\right)$:
$ET_{i,j} = \frac{r_{i1}}{R_{j1}}$.

\subsubsection{Transmission time} of a message $m_{ij} \in M$ of size $SZ_{ij} $ between two devices $d_{i}$ and $d_{j}$ is:
$T_{ij}= LAT_{ij} + \frac{SZ_{ij} }{BW_{ij}}$,
where $LAT_{ij}$ is the latency and $BW_{ij}$ is the bandwidth of a network connection $n_{ij} \in N$.

\subsubsection{Response time} of a service $s_i \in \mathcal{S}$ running on the device $d_j=\mu\left(s_i\right)$ is the sum between the maximum response time $RT_{pq}$ of its predecessors $s_p$, including its request message transmission time $T_{qj}$, where $\mu\left(s_p\right)=d_q$ (or $T_{uj}$, if initial message request), and its execution time $ET_{ij}$:
\[RT_{i,j} = \left\{
\begin{array}{@{}l@{}l@{}}
T_{uj} + ET_{ij}, & \exists m_{ui} \in M;\\
\underset{m_{pi} \in M}{\max}\left\{RT_{pq}+T_{qj}\right\}+ET_{ij}, & \exists m_{pi} \in M \land s_p\in\mathcal{S}.
\end{array}\right.\]

\subsubsection{Application response time} is the maximum response time of all its services $s_i \in S$ placed on devices $d_j=\mu\left(s_i\right)$:
\[RT_A = \underset{\forall s_{i} \in S}{\max}\left\{RT_{ij}\right\}.\]

\subsubsection{Deadline fulfilment} requires that the response time of the application placement satisfies the deadline $\theta$: $RT_A < \theta$.

\section{Fog multilayer partitioning}\label{sec:partition}
This section describes the multilayer Fog partitioning architecture, its design phases and the corresponding algorithm.

\subsection{Architecture design}
\begin{figure}[t]
	\centering
\includegraphics[width= \columnwidth]{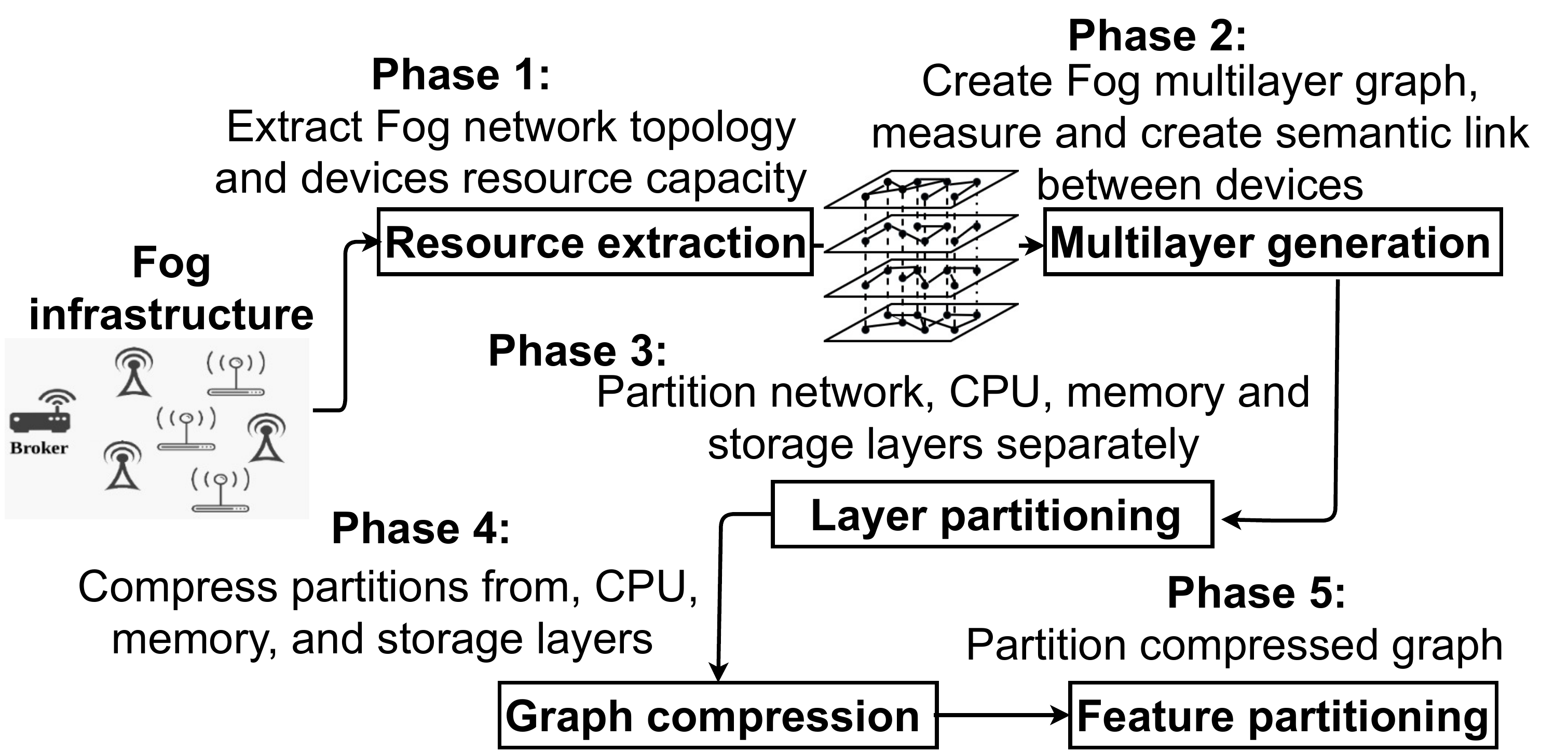}
	\caption{Fog multilayer partitioning architecture workflow.}
 \label{fig.arc}
\end{figure}

Figure~\ref{fig.arc} depicts the architecture design for the Fog multilayer resource partitioning in five phases: resource extraction, multilayer generation, layer partitioning, graph compression and feature partitioning.

\subsubsection{Resource extraction} identifies the infrastructure characteristics based upon different resource types (e.g. $R_{i1}, R_{i2}, R_{i3}$ representing speed of CPU cores in MIPS, memory size, and storage size) and reveals complete information about the Fog platform required for the correct application placement.

\subsubsection{Multilayer graph generation} models the Fog infrastructure as a multilayer graph in three steps:

\paragraph{Placement of Fog devices} in four layers based on their topological and semantic features.

\paragraph{Similarity score between Fog devices} based on different resource types.

\paragraph{Intra-layer Fog device linking} based on topological features and similarity scores, as defined in Section~\ref{subsec:fog-model}.

\subsubsection{Layer partitioning} splits each Fog multilayer graph layer $l \in L$ in a set $\mathcal{P}(l)$ of disjoint partitions that cluster the Fog devices based on their resource types (see Section~\ref{subsec:fog-model}).
\begin{enumerate}[font=\emph,wide]
    \item[(i)] \emph{Network layer $l_0$ partitioning} clusters the highly interconnected Fog devices based on their network connections $N$.
    \item[(ii)--(iv)] \emph{CPU $l_1$, memory $l_2$, and storage $l_3$ layer partitioning} cluster Fog devices with similar CPU speed ($R_{i1}$), memory size ($R_{i2}$), and storage size ($R_{i3}$).
\end{enumerate}

\subsubsection{Graph compression} groups the disjoint
CPU, memory and storage layer partitions in a high-level compressed graph representation associated to a similar resource type.

\subsubsection{Feature partitioning} splits the compressed graph in disjoint partitions such that each feature partition is a cluster of similar Fog devices across overlapping resource types.
\begin{figure*}[t]
\centering
	\begin{subfigure}[b]{0.25\textwidth}
		\includegraphics[width=\linewidth]{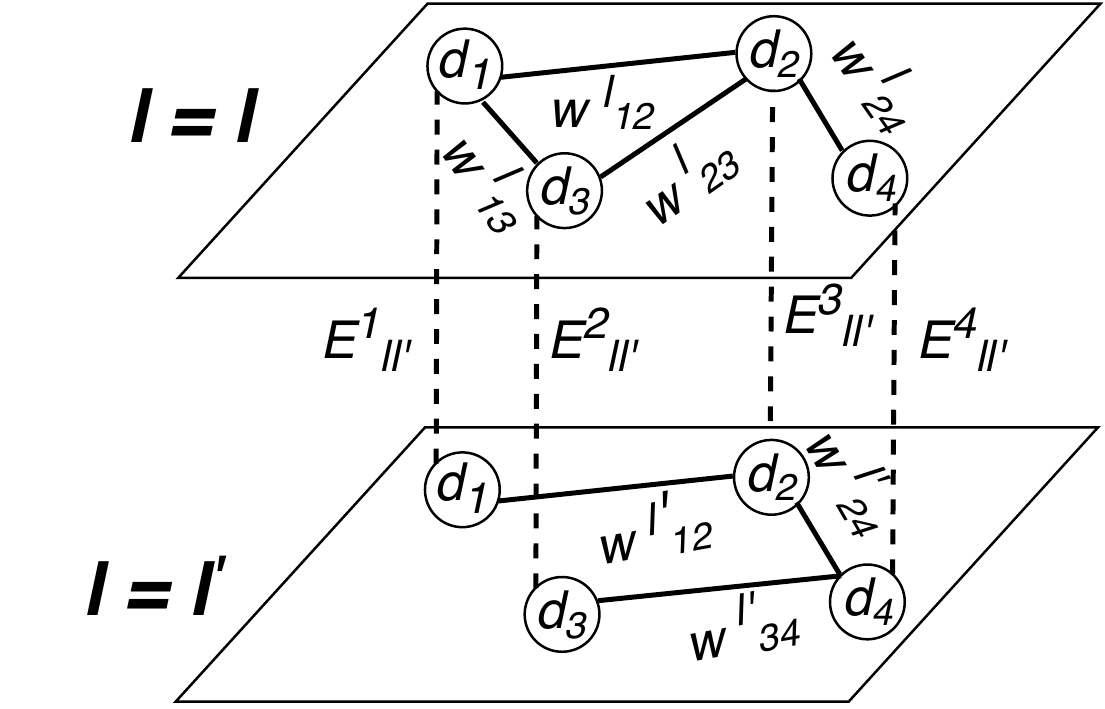}
	 	\caption{Multilayer graph example.}
		\label{fig:multiG}
	\end{subfigure}
	\begin{subfigure}[b]{0.73\textwidth}
	  \includegraphics[width=\linewidth]{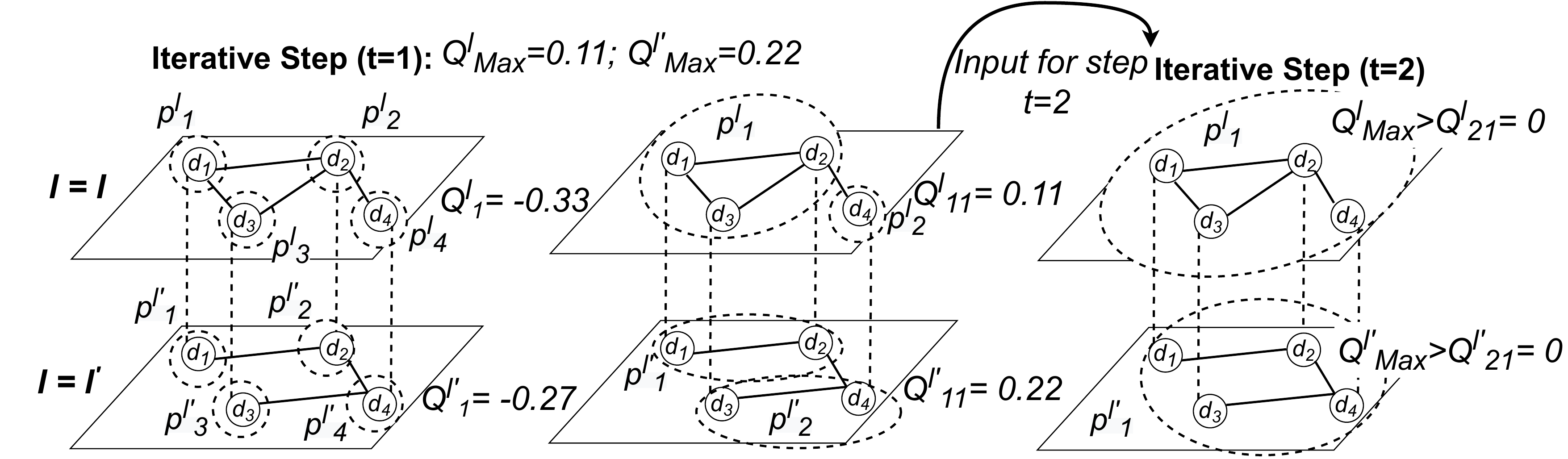}
		\centering
  \caption{Modularity ($Q$) computation at step $t=1$ and $t=2$ for the mutilayer graph from Figure~\ref{fig:multiG}.}
		\label{fig:modularity1}
	\end{subfigure}
	\begin{subfigure}[b]{0.23\textwidth}
   \centering
	  \includegraphics[width=\linewidth]{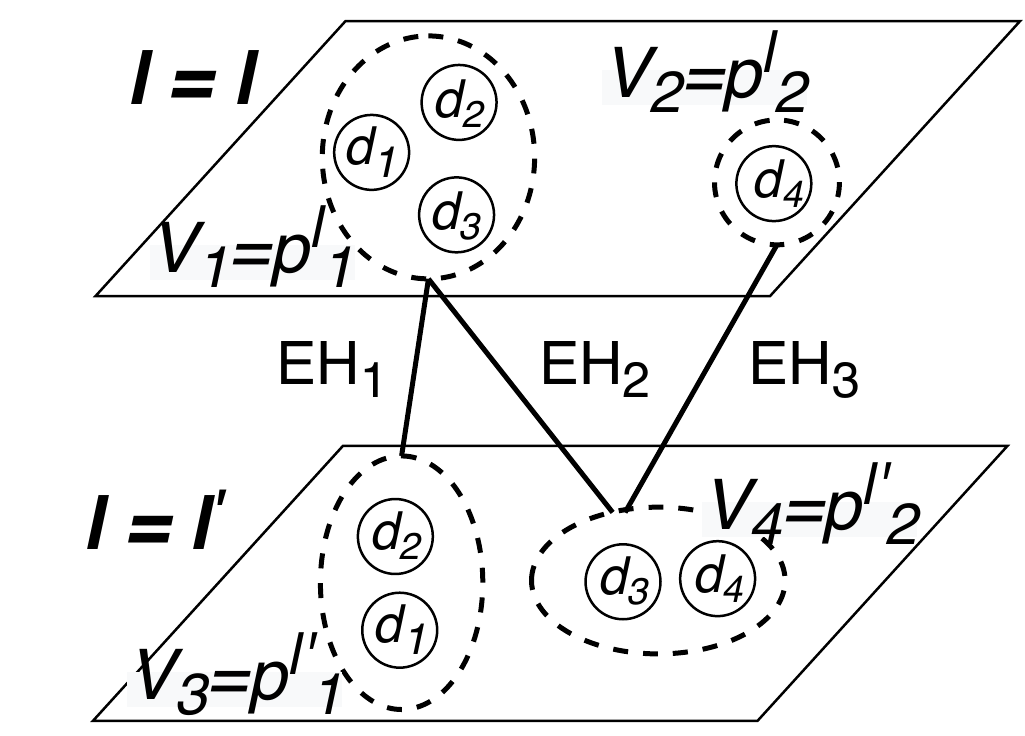}
		\caption{Compressed graph ($GH$) for the partitions in Figure~\ref{fig:modularity1}.}
		\label{fig:modularity3}
		\end{subfigure}
	\begin{subfigure}[b]{0.76\textwidth}
  \centering
	  \includegraphics[width=\linewidth]{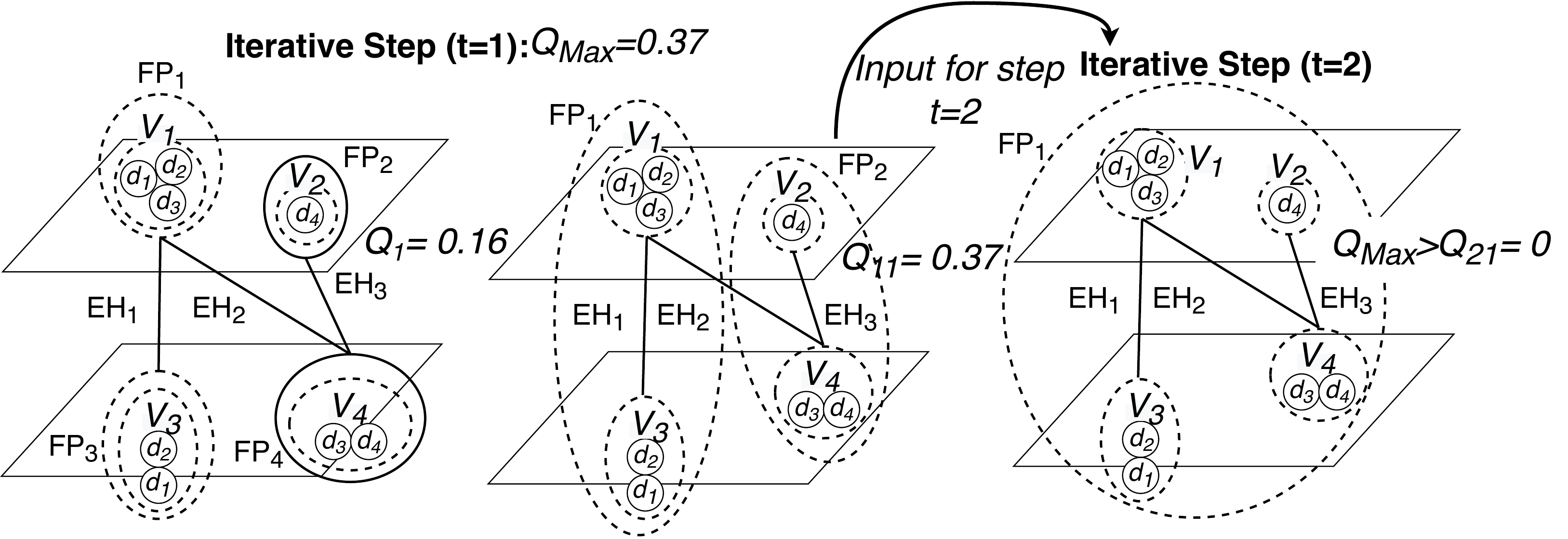}
		\caption{Feature partition ($FP$) for the compressed graph ($GH$) from Figure~\ref{fig:modularity3}.}
		\label{fig:modularity4}
	\end{subfigure}
		\caption{Fog multilayer graph, modularity computation, and Fog multilayer partitioning example.}
		\label{fig:multiGGG}
\end{figure*}
\subsection{Modularity background}
We use the \emph{modularity}~\cite{mucha2010community} metric $Q \in [-1,1]$ to measure the connectivity strength of Fog multilayer graph partitions:
\begin{multline*}
Q = \frac{1}{2W}\cdot\sum\limits_{l \in L}\sum\limits_{l^\prime \in L}\sum\limits_{d_i \in l}\sum\limits_{d_j \in l}\\
\left\{\left(w_{ij}^{(l)}-\frac{{\sigma^{l}_{i}}\cdot{\sigma^{l}_{j}}}{2W_{l}}\right)\cdot\delta_{ll^\prime}+\Delta_{ij}\cdot E^{j}_{ll^\prime}\right\}\cdot\lambda_{ij}, \mathrm{where:}
\end{multline*}
\begin{itemize}[align=left, leftmargin=*]
    \item $\sigma^{(l)}_{i}= \sum\limits_{d_j \in l} w_{ij}^{(l)}$ is the connectivity strength of the Fog device $d_i$ with the other devices in layer $l$;
    \item $\sigma^{(l)}_{j}= \sum\limits_{d_i \in L}^{n}w_{ij}^{(l)}$ is the connectivity strength of the Fog device $d_j$ with the other devices in layer $l$;
    \item $W_l= \sum\limits_{i=0}^{D}\sum\limits_{j=0}^{D}w_{ij}^{(l)}$ is the total sum of the link weights between $D$ Fog devices in each layer $l\in L$;
    \item $W=\sum\limits_{l\in L} {W_l}$ is the total sum of the link weights between Fog devices, $\forall l \in L$;
    \item $E^{(j)}_{ll^\prime}$ indicates the number of interlayer edges of the Fog device $d_j$ from layer $l$ to layer $l^\prime$;
    \item $\delta_{ll^\prime}$ is equal to 1 if $l=l^\prime$ and 0 otherwise;
    \item  $\Delta_{ij}$ is equal to 1 if $i=j$ and 0 otherwise; 
    \item $\lambda_{ij}$ is equal to 1 if device $d_i$ and $d_j$ belong to the same partition, otherwise 0.
\end{itemize}

\paragraph{$Q\leq 0$} represents low quality partitions of disassortative Fog devices with sparse connections among them.

\paragraph{$Q>0$} represents high quality topological partitions with better connectivity strength among densely-connected Fog devices. Hence, the goal is to find a set of partitions in a multilayer graph with highest modularity ($Q_G\to1$).

\subsection{Layer partitioning}\label{subsection:layerpar}
We apply the Louvain clustering technique~\cite{blondel2008fast} that utilizes the modularity metric~\cite{Newman8577} to obtain high quality partitions with densely connected Fog devices in each layer. We define the modularity of the obtained partitions as the difference between the number of edges within partitions and the expected number of edges over all pair of Fog devices. The Louvain algorithm
applies two phases in each iterative step $t>0$ until achieving partitions with the maximum modularity.

\subsubsection{Phase $1$} considers first each Fog device $d_i$ in layer $l$ as a single partition and calculates its modularity $Q_i$. Afterwards, it considers all neighbouring devices $d_j$ of $d_i$ (i.e. $\left(d_i,d_j\right) \in E_{ll}$ and $\left(d_j,d_i\right) \in E_{ll}$) and calculates the  modularity of the new possible partition $Q_{ij}$. If the gain in modularity is positive (i.e. $Q_{ij}-Q_i>0$), we place $d_i$ and $d_j$ in the same partition. We repeat this step sequentially for all Fog devices in layer $l$ until no further modularity gain is possible.

\subsubsection{Phase $2$} takes the set of partitions in each layer $l$ from the first phase and considers each partition as a node to build a new network in each layer. The links between the Fog devices in the same partition represent self-loops, while the links between Fog devices across partitions denote edges between nodes. The new network and the maximum modularity act as input to the next iterative step $t+1$ starting with phase $1$.

\subsubsection{Example} Figure~\ref{fig:multiG} shows a weighted multilayer graph $G = \left(D, \mathcal{E}, L\right)$ with two layers $L=\left\{l,l^\prime\right\}$. Each layer has four devices $D=\left\{d_1, d_2, d_3, d_4\right\}$ and the following intra-layer sets of edges: $E_{ll}=\left\{ \left(d_1,d_2\right),\left(d_1,d_3\right),\left(d_2,d_3\right)\right\}$ and $E_{l^{\prime}l^{\prime}}=\left\{\left(d_1,d_2\right),\left(d_3,d_4\right)\right\}$. The Louvain technique requires two iterative steps to find the partitions with the maximum modularity in each layer in this example.

\paragraph{Step $t=1$} shown in Figure~\ref{fig:modularity1} creates first a set of four partitions $\mathcal{P}(l)=\left\{p_{1}, p_{2}, p_{3}, p_{4}\right\}$ in layer $l$ and another $\mathcal{P}\left(l^{\prime}\right)=\left\{p^{\prime}_{1}, p^{\prime}_{2}, p^{\prime}_{3}, p^{\prime}_{4}\right\}$ in layer $l^{\prime}$. Each partition in $\mathcal{P}(l)$ and $\mathcal{P}\left(l^{\prime}\right)$ consists of a single Fog device ($d_1$, $d_2$, $d_3$ and $d_4$) with the modularities $Q_{\max}=-0.33$ and $Q^{\prime}_{\max}=-0.27$. Afterwards, it considers all neighboring devices to obtain two new partition sets $\mathcal{P}(l)=\left\{p_{1}, p_{2}\right\}$ and $\mathcal{P}\left(l^{\prime}\right)=\left\{p^{\prime}_{1}, p^{\prime}_{2}\right\}$ with a positive modularity gain (i.e. $Q_{1}=0.11>Q_{\max}$, \mbox{$Q^{\prime}_{1}=0.22>Q^{\prime}_{\max}$}), indicating a better connectivity strength of the devices in each layered partition. We therefore select the partitions $\mathcal{P}(l)$ and $\mathcal{P}\left(l^{\prime}\right)$, update the maximum modularities (i.e. $Q_{\max}=Q_1$, $Q^{\prime}_{\max}=Q^{\prime}_1$), and consider the partitions in $p_1, p_2, p^\prime_1, p^\prime_2$ as nodes in a new network over the two layers.

\paragraph{Step $t=2$} displayed in Figure~\ref{fig:modularity1} starts from the partitions $p_1, p_2, p^\prime_1, p^\prime_2$ with the maximum modularities $Q_{\max}$ and $Q^{\prime}_{\max}$ obtained in the step $t=1$, and considers the neighboring nodes to obtain the single partitions in each layer: $\mathcal{P}(l)=\left\{p_{1}\right\}$ and $\mathcal{P}\left(l^{\prime}\right)=\left\{p^{\prime}_{1}\right\}$ with the modularities $Q_{2}=Q^{\prime}_{2}=0$. As the maximum modularities from the step $t=1$ are positive for both layers, we consider them as the highly connected topological output.

\subsection{Graph compression}
\label{subsec:graph-compress}
Graph compression shrinks the disjoint partitions from the CPU, memory and storage layers and provides a high-level intermediate representation of partitions associated to similar resource types. The compressed graph merges all the similar resources inside a partition in a single node with an average capacity. This automates the computation of overlapping partitions without detailed analysis of individual Fog devices.

A \emph{compressed graph} $G_P = \left(V_P, E_P\right)$ corresponding to a multilayer graph $G=(D,\mathcal{E},L)$ consists of two sets:

\paragraph{Layer partition set} is the union of the partitions in the CPU ($l_1$), memory ($l_2$), and storage ($l_3$) layers: $V_P=\bigcup\limits_{l \in L}{\mathcal{P}(l)}$.

\paragraph{Inter-layer partition edges} represent connections between a partition $p$ in layer $l$ and a partition $p^\prime$ in layer $l^\prime \neq l$, such that there is at least one inter-layer edge $\left(d_i,d_j\right)\in E_{ll^\prime}$ in the original graph $G$ between a device $d_i\in p \in \mathcal{P}(l)$ and a device $d_j \in p^\prime \in \mathcal{P}\left(l^{\prime}\right)$:
\begin{multline*}
E_P = \left\{\left.\left(p,p^\prime\right) \in \mathcal{P}(l) \times \mathcal{P}\left(l^\prime\right)\ \right|\ \forall l \neq l^\prime \in L\right.\\ \left.\land \exists \left(d_i, d_j\right) \in E_{ll^\prime} \land d_i \in p \land d_j \in p^\prime\right\}.
\end{multline*}

\subsubsection{Example} Figure~\ref{fig:modularity3} illustrates a compressed graph that represents the four partitions in the $l$ and $l^{\prime}$ layers in Figure~\ref{fig:modularity1} as nodes: $V_P = \left\{p_1, p_2, p^\prime_1, p^\prime_2\right\}$. Similar to the nodes, we compress the edges between two partitions into: $E_p=\left\{ \left(p_1,p^\prime_1\right),\left(p_1,p^\prime_2\right),\left(p_2,p^\prime_2\right)\right\}$.

\subsection{Feature partitioning} \label{subsection:Featurepar}
We define a \emph{feature} of a layer partition $p \in V_P$ as a triplet $F_p = \left(R_{p1}, R_{p2}, R_{p3}\right)$ with average CPU speed, memory and storage sizes across all Fog devices in $p$.

Feature partitioning splits a compressed graph $G_P = \left(V_P, E_P\right)$ in a set $\mathcal{P}\left(G_P\right)$ of disjoint \emph{feature partitions} (exhibiting similar features) by applying $t$ iterative steps of the Louvain clustering technique as in the layer partitioning to achieve a maximum modularity (see Section~\ref{subsection:layerpar}).

\subsubsection{Example}
Figure~\ref{fig:modularity4} shows the two iterative steps of feature partitioning of the compressed graph $G_P$ in Figure~\ref{fig:modularity3}.

\paragraph{Step 1} initially creates a set of four feature partitions $\mathcal{P}\left(G_P\right)=\left\{FP_{1}, FP_{2}, FP_{3}, FP_{4}\right\}$ with the modularity $Q_{\max}=0.16$, where each feature partition $FP_{k}$ consists of a single layer partition (i.e. $p_1$, $p_2$, $p^\prime_3$ and $p^\prime_4$). Afterwards, it considers neighbouring partitions to obtain a new feature partition set $\mathcal{P}\left(G_P\right)=\left\{FP_{1}, FP_{2}\right\}$ with a positive modularity gain $Q=0.37>Q_{\max}$, which becomes the maximum modularity $Q_{\max}=Q$ at this step.

\paragraph{Step 2} starts from the feature partition set $\mathcal{P}\left(G_P\right)$ with the maximum modularity from the first step and considers each feature partition $FP_{k} \in \mathcal{P}\left(G_P\right)$ as a node of the new network (similar to layer partitioning). Afterwards, it iteratively checks the neighbouring feature partitions of each fearture partition $FP_{k} \in \mathcal{P}\left(G_P\right)$ and obtains a single partition $FP_{1}$ with a lower modularity $Q=0$. Hence, we select the feature partition set $\mathcal{P}\left(G_P\right)=\left\{FP_{1}, FP_{2}\right\}$ with the maximum modularity from the first step as the final output.
\subsection{Multilayer resource partitioning algorithm}
The \emph{multilayer resource partitioning algorithm} clusters Fog devices based on their network connections, CPU, memory, and storage resource characteristics. Algorithm~\ref{alg:partition} receives a Fog multilayer graph with four layers $L$, a set of Fog devices $D$ and their underlying CPU, memory, and storage resources, and the inter- and intra-layer edges $\mathcal{E}$ as input. Initially, lines~\ref{alg1:line1}--\ref{alg1:line2} initialize five empty sets corresponding to the partitions in the network ($l_0$), CPU ($l_1$), memory ($l_2$) and storage ($l_3$) layers, as well as the feature partitions. Thereafter, line~\ref{alg1:line3} performs the network layer partitioning $\mathcal{P}\left(l_{0}\right)$ that clusters densely connected Fog devices in the same partition. Similarly, lines~\ref{alg1:line4}--\ref{alg1:line6} partition the Fog devices in the CPU, memory, and storage layers and store them in $\mathcal{P}\left(l_{1}\right)$, $\mathcal{P}\left(l_{2}\right)$, and $\mathcal{P}\left(l_{3}\right)$ (see Section~\ref{subsection:layerpar}). Line~\ref{alg1:line7} creates a compressed graph $G_P\left(V_P,E_P\right)$ using inter-layer edges between the CPU, memory, and storage layers, where $V_P=\left\{\mathcal{P}\left(l_{1}\right),\mathcal{P}\left(l_{2}\right),\mathcal{P}\left(l_{2}\right)\right\}$ (see Section~\ref{subsec:graph-compress}). Afterwards, lines~\ref{alg1:line9}--\ref{alg1:line19} compute the feature triplet $F_P$ of each partition $p\in V_P$ as the average CPU speed, memory, and storage sizes of their Fog devices $d_i \in p$. Line~\ref{alg1:line20} performs feature partitioning of the compressed graph partitions with similar features, as presented in Section~\ref{subsection:Featurepar}. Finally, line~\ref{alg1:line21} returns the feature partition set $\mathcal{P}\left(G_P\right)$ and the set of partitions in the network layer $\mathcal{P}\left(l_{0}\right)$.
\begin{algorithm}[t]
  \scriptsize
  \SetKwInOut{Input}{Input}
  \SetKwInOut{Output}{Output}
  \SetKwFunction{createMultilayerGraph}{createMultilayerGraph}
  \SetKwFunction{layerPartitioning}{layerPartition}
  \SetKwFunction{graphCompression}{graphCompress}
  \SetKwFunction{featurePartitioning}{featurePartition}
  \Input{
    $G = (D, \mathcal{E}, L)$: Fog multilayer graph\\
    $L=\left\{l_0, l_1, l_2, l_3\right\}$: Fog layers\\
    $D = \left\{d_i | d_i = \left(R_{i1}, R_{i2}, R_{i3}\right)\right\}$: set of Fog devices\\
    $\mathcal{E} = \left\{ E_{ll^\prime}\ |\ \forall l, l^\prime \in [0, L)\right\}$: inter- and intra-layer edges\\
  }
  \Output{
    $\mathcal{P}\left(G_P\right)$: feature partition set\\
    $\mathcal{P}\left(l_{0}\right)$: network layer partition set\\
  }
  $\mathcal{P}\left(l_{0}\right)\gets \emptyset$; $\mathcal{P}\left(l_{1}\right)\gets \emptyset$; $\mathcal{P}\left(l_{2}\right)\gets \emptyset$; $\mathcal{P}\left(l_{3}\right)\gets \emptyset$\\ \label{alg1:line1}
  $\mathcal{P}\left(G_P\right) \gets \emptyset$\\ \label{alg1:line2}
  $\mathcal{P}\left(l_{0}\right)\gets$ \layerPartitioning{$D,E_{l_{0}l_{0}},l_0$}\\ \label{alg1:line3}
  $\mathcal{P}\left(l_{1}\right)\gets$ \layerPartitioning{$D,E_{l_{1}l_{1}},l_1$}\\ \label{alg1:line4}
  $\mathcal{P}\left(l_{2}\right)\gets$ \layerPartitioning{$D,E_{l_{2}l_{2}},l_2$}\\ \label{alg1:line5}
  $\mathcal{P}\left(l_{3}\right)\gets$ \layerPartitioning{$D,E_{l_{3}l_{3}},l_3$}\\ \label{alg1:line6}
  $G\left(V_P,E_P\right) \gets$ \graphCompression{$\mathcal{P}\left(l_{1}\right),\mathcal{P}\left(l_{2}\right),\mathcal{P}\left(l_{3}\right),E_{l_{1}l_{2}},E_{l_{1}l_{3}},E_{l_{2}l_{3}}$}\\ \label{alg1:line7}
  $fList \gets \emptyset$;\\ \label{alg1:line8}
  \ForAll{$p \in V_P$}{\label{alg1:line9}
        $R_{p1}\gets\underset{\forall d_i \in p} {avg}\left\{R_{i1}\right\}$;
        $R_{p2}\gets\underset{\forall d_i \in p} {avg}\left\{R_{i2}\right\}$; $R_{p3}\gets\underset{\forall d_i \in p} {avg}\left\{R_{i3}\right\}$\\
        $F_p \gets \left(R_{p1}, R_{p2}, R_{p3} \right)$\\
        $fList \gets fList \cup F_p$\\
     }\label{alg1:line19}
     $\mathcal{P}\left(G_P\right) \gets $ \featurePartitioning{$G_P\left(V_P, E_P\right), fList$}\\ \label{alg1:line20}
  \Return $(\mathcal{P}\left(G_P\right), P\left(l_{0}\right))$\label{alg:par:return} \label{alg1:line21}
  \caption{Multilayer resource partitioning.}
  \label{alg:partition}
\end{algorithm}

\section{Multilayer Fog application placement}
\label{sec:placement}
This section describes the application placement workflow in multilayer Fog partitions, its design, and the underlying feature partition selection and service placement algorithms.

\subsection{Multilayer Fog placement workflow} 
The multilayer Fog placement of application requests $A=\left(\mathcal{S},M,\theta,u\right)$ coming from end-users has two phases.
\paragraph{Feature partition selection} maps each service $s\in\mathcal{S}$ of the requested application to an appropriate feature partition $FP_{k}$ composed of layer partitions $p \in V_P$ with the feature $F_p$ similar to the resource demand of the service $s$.

\paragraph{Service placement} allocates a Fog device $d = \mu\left(s\right)$ to each service $s\in\mathcal{S}$ in the selected feature partitions, such that its selected Fog devices exist in the same network layer partition. This enables placing interrelated services of the same application across highly connected Fog devices.

\subsection{Service fitness}
\label{subsec:fitness}
An application $A = \left(\mathcal{S}, M, \theta, u\right)$ has a set of resource requirements for each service $s_i=\left(r_{i1},r_{i2},r_{i3}\right)$ in terms of CPU speed, memory size and storage capacity required for a successful execution. Hence, it is imperative to place each service $s_i \in \mathcal{S}$ across the Fog devices of a feature partition $d \in FP_k$  composed of layer partitions $p \in FP_k$ with the feature $F_p$ that satisfies these requirements. Additionally, the service placement requires the Fog devices in close proximity of the user location $u$ to simultaneously satisfy application latency and deadline constraints $\theta$.

We define the \emph{fitness} of a service $s_i \in \mathcal{S}$ for a feature partition $FP_k$ requested by an user $u$ as follows:
\begin{multline*}
Fit\left(FP_k,s_i,u\right) = \alpha \cdot  
\underset{\forall p \in FP_k}{\max}
\left\{Sim\left(p, s_i\right)\right\}
+\\ + \beta
\cdot \left(\frac{1}{1+\min\limits_{\forall d_j \in FP_k} \left\{T_{uj}\right\}}\right),\mathrm{where:}
\end{multline*}

\paragraph{$\underset{\forall p \in FP_k}{\max} \left\{Sim\left(p,s_i\right)\right\}$} is the maximum similarity between the partition with the feature $F_p = \left(R_{p1},R_{p2},R_{p3}\right)$ in feature partition $FP_k$  and the resource demand of the service $s_i = \left(r_{i1},r_{i2},r_{i3}\right)$, computed using the Euclidean distance between the feature $F_p$ and the resource demand $s_i$ in all three dimensions, normalized in the $[0,1]$ interval;

\paragraph{$\underset{\forall d_j \in FP_k} {\min}\left\{T_{uj}\right\}$} is the minimum transmission time of a message $M_{uk}$ of size $SZ_{uk}$ between the source (user) $u$ and the destination Fog devices $d_j$ in the feature partition $FP_{k}$;

\paragraph{$\alpha, \beta$} are weighting factors for the similarity and transmission times in the service fitness calculation.

\subsection{Feature partition selection algorithm} 
The \emph{feature partition selection algorithm} places the services of requested applications to appropriate feature partitions composed of Fog devices that satisfy the application deadline $\theta$ and individual service resource demands.

Algorithm~\ref{alg:Placepar} takes as input a set of requested applications $AS$ (including their services, resource demands and deadlines), a set of Fog devices $D$, the feature partition set $\mathcal{P}\left(G_P\right)$, the network partition set $P\left(l_{0}\right)$ (both computed by Algorithm~\ref{alg:partition}) and the message transmission times $\mathcal{T}$ between user and Fog devices. Firstly, line~\ref{alg2:line1} sorts the applications to prioritize the placement of those with the lowest deadline. Lines~\ref{alg2:line2}--\ref{alg2:line4} select the appropriate feature partitions for all applications $A \in AS$ by placing all their services $\mathcal{S} \in A$ on Fog devices that satisfy their resource demands in the proximity of the requesting users. The algorithm returns an array of placement functions $\mu{}List[AS]$ in line~\ref{alg2:line3}. $\forall A \in AS$.

To select the feature partition (line~\ref{alg2:line7}) that satisfies the resource demand of each service $s_{i} \in S$  (line~\ref{alg2:line8}), lines~\ref{alg2:line10}--\ref{alg2:line13} iterate through each feature partition $FP\in \mathcal{P}\left(G_P\right)$), calculate its fitness to service $s \in \mathcal{S}$ (see Section~\ref{subsec:fitness}), and insert it in a $fpRank$ list in descending fitness order (line~\ref{alg2:line11}). Line~\ref{alg2:line12} sorts the Fog devices $d \in FP$ in ascending order based on the transmission time $T_{ui}$ and stores them in a two dimensional array $dMatrix$ to enable their placement in user proximity. Line~\ref{alg2:line14} invokes a service placement function (see Algorithm~\ref{alg3:applacement}) that maps each service $s \in \mathcal{S}$ of the same application across the selected feature partitions with Fog devices in the same network layer partition. The function returns the Fog device $\mu\left(s\right)$, which updates the application placement function $\mu$ returned by the algorithm in line~\ref{alg2:line17}.

\begin{algorithm}[t]
     \scriptsize
     \SetKwInOut{Input}{Input}
     \SetKwInOut{Output}{Output}
     \SetKwFunction{sortApp}{sortApp}\SetKwFunction{Fit}{Fit}\SetKwFunction{featurePlacement}{placeService}\SetKwFunction{selectFP}{selectFP}\SetKwFunction{insert}{insert}\SetKwFunction{sortDev}{sortDev}
     \SetKwProg{Fn}{Function}{:}{}
     \Input{
        $AS = \left\{A | A = (\mathcal{S}, M, \theta, u)\right\}$: set of applications;\\
        $D = \left\{d_i | d_i= \left(R_{i1}, R_{i2}, R_{i3}\right)\right\}$: set of Fog devices;\\
        $\mathcal{P}\left(G_P\right)$: feature partition set;\\
        $\mathcal{P}\left(l_{0}\right)$: network layer partition set; \\
        $\mathcal{T} =\left\{\left.T_{ui} \right| \left(u, d_i\right) \in \mathcal{U} \times D\right\}$: message $M_{ui}$ transmission times;\\
     }
     \Output{
         $\mu{}List[AS]$: array of  $\mu{}List[A]$
         service placements, $\forall A\in AS$;
     }
     $AS \gets$\sortApp{$AS,\theta$}\\ \label{alg2:line1}
     \ForAll{$A=\left(\mathcal{S},M,\theta,u\right) \in AS$}{\label{alg2:line2}
       $\mu{}List[A] \gets$ \selectFP{$\mathcal{S},u$}\\ \label{alg2:line3}
     }\label{alg2:line4}
     \Return $\mu{}List$\\\label{alg2:line5}
\;\label{alg2:line6}
\Fn{\selectFP{$\mathcal{S},u$}}{\label{alg2:line7}
        \ForAll{$s \in \mathcal{S}$}{\label{alg2:line8}
            $fpRank \gets \emptyset$\\ \label{alg2:line9}
            \ForAll{$FP\in\mathcal{P}\left(G_P\right)$}{\label{alg2:line10}
                $fpRank \gets$ \insert{$fpRank$, \Fit{$FP,s,u$}}\\ \label{alg2:line11}
                $dMatrix[FP]\gets$ \sortDev{$FP,T$}\\ \label{alg2:line12}
            }\label{alg2:line13}
            $\mu\left(s\right)\gets$ \featurePlacement{$P\left(l_{0}\right), fpRank, dMatrix, s$}\\ \label{alg2:line14}
        }\label{alg2:line15}
     }\label{alg2:line16}
     \Return $\mu$\label{alg2:line17}
     \caption{Feature partition selection.}
     \label{alg:Placepar}
\end{algorithm}

\subsection{Service placement algorithm} 
Algorithm~\ref{alg3:applacement} places a service $s$ on a Fog device $\mu(s)$ in the same network layer partition as all other services of the same application. The input arguments to Algorithm~\ref{alg3:applacement} are
\begin{enumerate*}
    \item a network layer partition set $P\left(l_{0}\right)$,
    \item a feature partition set $\mathcal{P}\left(G_P\right)$ ranked based on fitness,
    \item a sorted list of Fog devices based on their transmission time in each feature partition $FP\in \mathcal{P}\left(G_P\right)$, and
    \item the placement service $s_i$ (line~\ref{alg3:line1}).
\end{enumerate*}
First, lines~\ref{alg3:line2}--\ref{alg3:line3} extract the network layer partition of the first service placement $\mu\left(s_1\right)$ in $p_1$ (if available).
Lines~\ref{alg3:line4}--\ref{alg3:line12} iterate through all the feature partitions $FP\in\mathcal{P}\left(G_P\right)$ in descending order of their fitness. Afterwards, line~\ref{alg3:line5} extracts the set of Fog devices $dList$ in each feature partition $FP$ sorted by the transmission times to the requesting user. To place the service $s_i$ onto a Fog device, lines~\ref{alg3:line6}--\ref{alg3:line11} iterate through each device $d_j\in dList$ and line~\ref{alg3:line7} extracts its network layer partition in $p_i$. If this partition is the same as $p_1$ and the device $d_j$ meets the resource constraints of the service $s_i$, line~\ref{alg3:line9} performs the placement. If no service placement on the same network partition is possible, line~\ref{alg3:line13} assigns an invalid device. Finally, lines~\ref{alg3:line10} and \ref{alg3:line14} return the service placement result.

\begin{algorithm}[t]
  \scriptsize
  \SetKwInOut{Input}{Input}
  \SetKwFunction{featurePlacement}{placeService}\SetKwFunction{sort}{sort}
  \SetKwFunction{getNetworkLayerPartition}{getNetPartition}
  \SetKwFunction{updateFogDeviceResource}{updateFogDeviceResource}
  \SetKwFunction{mutlilayerPlacement}{mutlilayerPlacement}
  \SetKwFunction{getsortedDevices}{getSortedDevices}
  \SetKwProg{Fn}{Function}{:}{}
  \Input{
    $\mathcal{P}\left(l_{0}\right)$: network layer partition set\\
    $\mathcal{P}\left(G_P\right)$: fitness-ranked feature partition set\\
    $dMatrix[\mathcal{P}\left(G_P\right)]$: sorted Fog devices, $\forall FP\in \mathcal{P}\left(G_P\right)$\\
    $s_i = \left\{r_{i1},r_{i2},r_{i3}\right\}$: service to place\\
  }
\Fn{\featurePlacement{$P\left(l_{0}\right), \mathcal{P}(G_P), dMatrix, s_i$}}{\label{alg3:line1}
  \uIf{$s_{i} \neq s_{1}$}{\label{alg3:line2}
            $p_{1} \gets$\getNetworkLayerPartition{$\mu\left(s_1\right), \mathcal{P}\left(l_{0}\right)$}\label{alg3:line3}\\
  }\
  \ForAll{$FP \in \mathcal{P}\left(G_P\right)$}{\label{alg3:line4}
    $dList \gets dMatrix[FP]$\\\label{alg3:line5}
    \ForAll{$d_j = \left(R_{j1},R_{j2},R_{j3}\right) \in dList$}{\label{alg3:line6}
            $p_{i} \gets$\getNetworkLayerPartition{$d, P\left(l_{0}\right)$}\label{alg3:line7}\\
        \uIf{$p_{1}=p_{i} \land \frac{r_{i1}}{R_{j1}} \leq \theta \land r_{i2} \leq R_{j2} \land r_{i3} \leq R_{j3}$}{\label{alg3:line8}
            $\mu\left(s_i\right) \gets d_j$\\ \label{alg3:line9}
            \Return $\mu(s_i)$\\ \label{alg3:line10}
        }
    }\label{alg3:line11}
  }\label{alg3:line12}
  $\mu(s_i) \gets \varnothing$\label{alg3:line13}\\
  \Return $\mu(s_i)$\label{alg3:line14}
}
\caption{Service placement.}
\label{alg3:applacement}
\end{algorithm}

\section{Evaluation}
\label{sec:eval}
We present first our experimental setup, then analyze the results of our multilayered partitioning method against two related resource~\cite{taneja2017resource} and availability-aware~\cite{lera2018availability} methods.

\subsection{Experimental setup}
We simulated a Fog infrastructure using the \texttt{YAFS}~\cite{lera2019yafs} simulator on an \texttt{Intel$^\circledR$ Core$^{(TM)}$ i7-8650U} server at \SI{1.90}{\giga\hertz} running Ubuntu 18.04 (\texttt{x86\_64}) operating system with \SI{16}{\giga\byte} of DDR4 RAM memory. We simulated the Fog infrastructure (e.g. devices, network topology), applications, and clients using similar configuration settings as two previous studies~\cite{lera2018availability,taneja2017resource} for a fair comparison.

\subsubsection{Fog infrastructure}
We simulated a Fog infrastructure as a bidirectional graph based on a Albert-Barbasi random network~\cite{barabasi2009scale} of $100$ devices, where $25$ devices with the lowest betweenness centrality~\cite{freeman1977set} represent gateways at the edge of the network. We represented the Cloud data center though an additional device with the highest betweenness centrality computed using the Python \texttt{networkx} module. We simulated resource characteristics (e.g. cores, CPU speed, memory, storage) of each Fog device using a uniform random distribution within the range specified in Table~\ref{tab:fogSetup}. Finally, we configured the bandwidth and latency across the Fog network as specified in Table~\ref{tab:netSetup}, similar to previous studies~\cite{lera2018availability,taneja2017resource}.

\begin{table}[t]
 \caption{Experimental setup.}
    \begin{subtable}[t]{0.4\columnwidth}
        \centering
        \caption{Fog device.}
        \begin{tabular}{|@{ }c@{ }|@{ }c@{ }|}
        \hline
        \emph{Parameters} & \emph{Range}\\
        \hline
        CPU cores & \SIrange{10}{25}{}\\
        CPU speed & \SIrange{20}{60}{MI\per\second}\\
        Memory size & \SIrange{10}{25}{\giga\byte}\\
        Storage size & \SIrange{10}{25}{\tera\byte}\\
        \hline
       \end{tabular}
       \label{tab:fogSetup}
    \end{subtable}
    \begin{subtable}[t]{0.5\columnwidth}
        \centering
         \caption{Application.}
        \begin{tabular}{|@{ }c@{ }|@{ }c@{ }|}
        \hline
        \emph{Parameters} & \emph{Value}\\
        \hline
        Services & \SIrange{2}{10}{}\\
        Deadline & \SIrange{300}{50000}{\milli\second}\\
        Memory size & \SIrange{1}{6}{\giga\byte}\\
        Storage size & \SIrange{1}{6}{\tera\byte}\\
        Message size & \SIrange{1500}{4500}{\kilo\byte}\\
        Workload & \SIrange{20}{60}{MI}\\
        \hline
        \end{tabular}
        \label{tab:appSetup}
     \end{subtable}
    \begin{subtable}[t]{0.49\columnwidth}
        \centering
        \caption{Network.}
        \begin{tabular}{|@{ }c@{ }|@{ }c@{ }|}
        \hline
        \emph{Parameters} & \emph{Value}\\
        \hline
        Latency & \SI{5}{\milli\second}\\
        Bandwidth & \SI{75000}{\byte\per\milli\second}\\
        \hline
       \end{tabular}
       \label{tab:netSetup}
    \end{subtable}
     \hfill
    \begin{subtable}[t]{0.49\columnwidth}
        \centering
        \caption{Client.}
        \begin{tabular}{|@{ }c@{ }|@{ }c@{ }|}
        \hline
        \emph{Parameters} & \emph{Value}\\
        \hline
        Request rate & \SI{1.557}{\milli\second}\\
        \hline
       \end{tabular}
       \label{tab:cliSetup}
    \end{subtable}
     \label{tab:expSetup}
\end{table}

\subsubsection{Applications}
We simulated an application $A$ as a directed graph using the Python \texttt{networkx.Gn\_Graph} module, where the nodes correspond to the application services. We generated a number of services for each application $A$ in the range between two and ten, and their resource requirements (i.e. CPU speed, number of cores, memory and storage size) using a uniform random distribution, as described in Table~\ref{tab:appSetup}. Additionally, we defined the dependencies between two services in terms of request messages of sizes between \SIrange{1500}{4500}{\kilo\byte}, which generated a workload in the destination service in the range \SIrange{20}{60}{MI}.

\subsubsection{Client}
We configured the client layer similar to previous studies~\cite{lera2018availability,taneja2017resource} such that end-users connected to gateway devices in Fog layer request random applications every \SI{1.557}{\milli\second}, as specified in Table~\ref{tab:cliSetup}. 

\subsubsection{Evaluation metrics}
We identified five performance objectives to optimize the multilayer application Fog placement $\mu$ for a set of applications $AS$.

\paragraph{Placement success rate}
is the ratio between the number of successfully placed services $\mathcal{S}_{\mu}$ and the total number of services $\mathcal{S}$ of all applications $A =(\mathcal{S},M,\theta,u) \in AS$:
$\frac{\sum_{A \in AS}{\left|\mathcal{S}_{\mu}\right|}}{\sum_{A \in AS}\left|\mathcal{S}\right|}$,
where $\mathcal{S}_{\mu} = \left\{s \in A | \mu\left(s\right) \neq \varnothing \right\}$ and $\left|\mathcal{S}\right|$ and $\left|\mathcal{S}_{\mu}\right|$ represent the cardinality of the two service sets.

\paragraph{Resource wastage} is the remaining percentage between total resource units consumed by the placed services to the total CPU, memory and disk resource units of the devices:
\[1-\frac{\sum\limits_{A \in AS}\sum\limits_{s_i \in \mathcal{S}_{\mu}}{\max\left\{\frac{r_{i1}}{CPU}, \frac{r_{i2}}{RAM},\frac{r_{i3}}{Disk}\right\}}}{\sum\limits_{d_j \in D}{\max\left\{\frac{R_{j1}}{CPU}, \frac{R_{j2}}{RAM},\frac{R_{j3}}{Disk}\right\}}}.
\]
We define a \emph{resource unit} as a triplet: $(CPU core,RAM,Disk)=(1,\SI{1}{\giga\byte},\SI{1}{\tera\byte})$. The resource units consumed by a service or device is the maximum of these three unit components.

\paragraph{Deadline satisfaction} represents the ratio between the total number of applications that fulfill their deadline $A_\theta=\left\{A \in AS | RT_A < \theta\right\}$ and the total number of applications $|AS|$ requested for placement: $\frac{\left|A_\theta\right|}{|AS|}$.

\paragraph{Hop distance} indicates the proximity of the placed services to the requesting users. A hop distance of $0$ indicates a service placement at the Fog gateway device. We compute a hop distance histogram across all application services, aiming to increase the number of placements at a low hop distance.

\subsubsection{Evaluation scenarios}\label{sec:evalS}
Similar to the study in~\cite{lera2018availability}, we divide our evaluation into two parts, summarized in Table~\ref{tab:eval-scenarios}.

\begin{table}[t]
\centering
\caption{Evaluation scenarios.}
\begin{tabular}{|@{ }c@{ }|@{ }c@{ }|@{ }c@{ }|@{ }c@{ }|@{ }c@{ }|@{ }c@{ }|@{ }c@{ }|}
\hline
\emph{Evaluation} & \emph{Scenario} & \emph{App.} & \emph{Service} & \emph{Users} & \emph{App.} & \emph{Service}\\
\emph &  &  &  &  & \emph{Requests} & \emph{Requests}\\
\hline
\emph{Service} & SMALL & 10 & 63 & 29 & 29 &204\\
\emph{placement} & MEDIUM & 20 & 129 & 65 & 65 & 440\\
& LARGE & 30 & 179 & 98 & 98 & 537\\
\hline
\emph{Deadline} & D-SMALL & 10 & 63 & 29 & 84429 & 573458 \\
\emph{fulfilment} & D-MEDIUM & 20 & 129 & 65 & 161597 & 933944 \\
& D-LARGE & 30 & 179 & 98 & 216038 & 1163571 \\
\hline
\end{tabular}
\label{tab:eval-scenarios}
\end{table}

\paragraph{Service placement} evaluates the placement success rate, resource wastage and hop distance for three placement scenarios (i.e. SMALL, MEDIUM, and LARGE) with different randomly generated application sets, number of services, users, application requests and corresponding service requests. Every user randomly requests one application for placement.

\paragraph{Deadline satisfaction} evaluates three scenarios (i.e. D-SMALL, D-MEDIUM, D-LARGE) in a reliable and faulty Fog infrastructure across a simulation period of \SI{2000}{\second}. 
We randomly introduced failures across the Fog devices every \SI{20}{\second} in the faulty simulation, similar to~\cite{lera2018availability}. All users select a random application every \SI{1.557}{\second} until the complete simulation.

\subsubsection{Related work} We compare our multilayer resource-aware partitioning approach with two state-of-the-art methods:

\paragraph{Availability-aware placement} with improved application deadline satisfaction in presence of device failures~\cite{lera2018availability}. 

\paragraph{Resource-aware placement} with optimized resource usage and reduced latency and application response time by applying fractional selectivity model~\cite{taneja2017resource}.
\subsection{Service placement}
\begin{wrapfigure}{R}{0pt}
  \centering
  \includegraphics[width=.5\columnwidth]{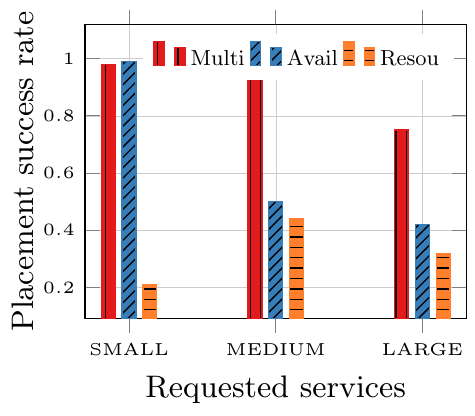}
	 \caption{Service placement.}
  \label{fig:placedser}
\end{wrapfigure}

Figure~\ref{fig:placedser} shows the placement success rate for the three evaluation scenarios.

\paragraph{SMALL} The multilayer and availability-aware methods performed similarly by placing $200$ and $202$ services with a similar success rate of $0.98$ and $0.99$, respectively. The resource-aware approach placed only $161$ services with a lower success rate of $0.78$.

\paragraph{MEDIUM and LARGE} The  availability-aware approach placed $221$ and $241$ services with a success ratio of $0.50$ and $0.44$, while the resource-aware approach placed only $188$ and $172$ services with a low placement ratio of $0.42$ and $0.32$, respectively. In contrast, the multilayer method outperformed them by placing $419$ and $407$ services with $0.95$, respectively  $0.75$ placement success rates.

\paragraph{Summary} The multilayer method is able to place more services with the increasing number of requests by mapping their individual resource demands to unique Fog device resource characteristics (i.e. CPU, RAM, storage). In contrast, the availability-aware approach prioritizes a set of dependent services by placing them onto the same Fog device if it satisfies their joint resource demands only. While resource-aware approach place dependent services across different Fog devices if they satisfy their joint bandwidth requirements.

\subsection{Resource wastage} 
\begin{figure}[t]
\centering
    \begin{subfigure}[t]{.48\columnwidth} 
    \centering
       \includegraphics[width=\linewidth]{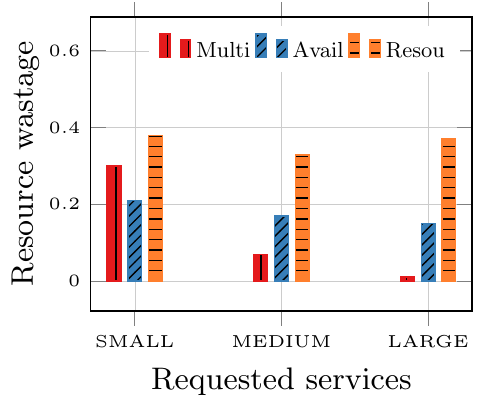}
       \vspace*{-0.6cm}
	    	\caption{Resource wastage.}
	   	\label{fig:ResWastage:a}
		\end{subfigure}
    \begin{subfigure}[t]{.5\columnwidth} 
    \centering
       \includegraphics[width=\linewidth]{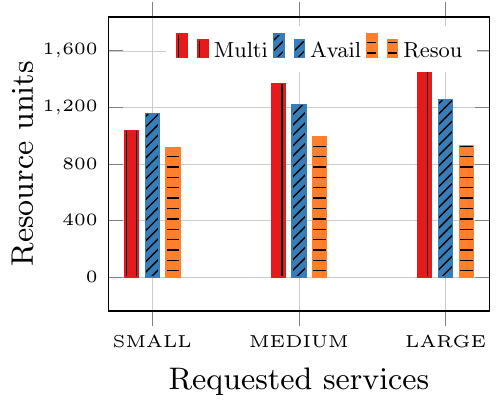}
       \vspace*{-0.6cm}
		   \caption{Resource unit consumption.}
		   \label{fig:ResWastage:b}
		\end{subfigure}
 		\caption{Resource wastage.}
		\label{fig:ResWastage}
		\vspace*{-0.6cm}
\end{figure}

Figure~\ref{fig:ResWastage:a} compares the resource wastage ratio for the multilayer, availability-aware and resource-aware placement approaches across the simulated Fog infrastructure ($100$ Fog devices) with $1483$ resource units. Figure~\ref{fig:ResWastage:b} explains the results through the resource units consumed by each placement.

\paragraph{SMALL} The multilayer method placed $200$ services with a resource wastage of $0.30$, while the availability-aware approach performed slightly better by placing $202$ services with a resource wastage of $0.21$. The resource-aware approach performed the worst by placing only $161$ services with a high resource wastage of $0.38$. The multilayer and availability-aware methods with similar placement success rate consumed $1036$, respectively $1157$ resource units. In comparison, the resource-aware approach consumed only $913$ resource units for a low placement success rate.

\paragraph{MEDIUM} The availability-aware and resource-aware approaches placed $221$ services with a resource wastage of $0.17$ and $0.33$, respectively. The multilayer method outperformed them and placed $419$ services with a low resource wastage of $0.07$. The multilayer method consumed $1367$ resource units with a higher placement success rate compared to the availability-aware and resource-aware approaches consuming only $1218$, respectively $989$ resource units.

\paragraph{LARGE} Similarly, the multilayer method performed the best by placing $407$ services across $100$ devices with a very low resource wastage of $0.011$. In contrast, the availability and resource-aware approaches placed $241$ and $172$ services with a high resource wastage of $0.15$, respectively $0.37$. The multilayer method consumed $1466$ resource units with increasing requests due to its higher placement success rate. In contrast, the availability-aware and resource-aware approaches consumed only $1257$, respectively $928$ resource units.

\paragraph{Summary}
The multilayer maximizes the placement success rate for increasing requests and thus, consume more resource units of the Fog infrastructure, and reduces the resource wastage upto $15$ times compared to availability-aware and upto $32$ times to resource-aware method.
\subsection{Hop distance}
\begin{figure}[t]
  \centering
  \begin{subfigure}[t]{0.39\columnwidth}
      \centering
      \includegraphics[width=\linewidth]{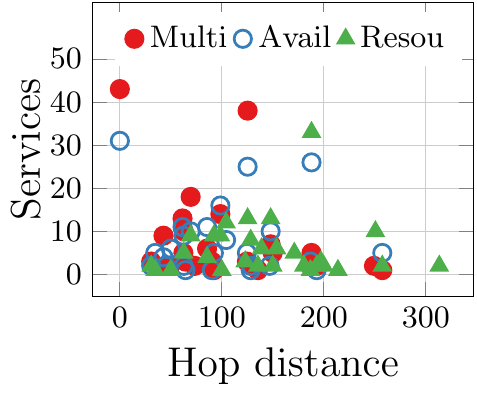}
      \vspace*{-0.6cm}
		  \caption{SMALL.}
		  \label{fig:hopDisN:b}
		\end{subfigure}
    \begin{subfigure}[t]{0.41\columnwidth}
      \centering
      \includegraphics[width=\linewidth]{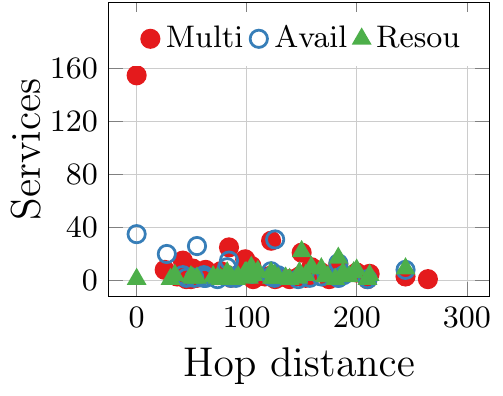}
      \vspace*{-0.6cm}
		   \caption{MEDIUM.}
		   \label{fig:hopDisN:d}
		\end{subfigure}
		\begin{subfigure}[t]{0.39\columnwidth}
		   \centering
      \includegraphics[width=\linewidth]{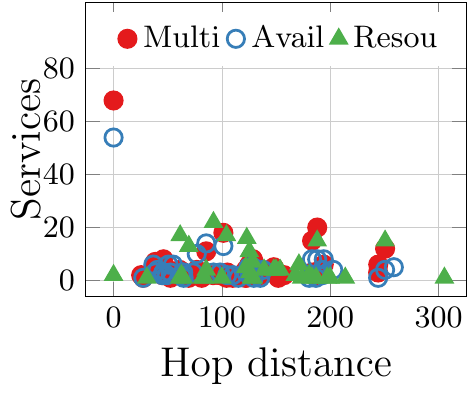}
      \vspace*{-0.6cm}
		   \caption{LARGE.}
		   \label{fig:hopDisN:e}
		\end{subfigure}
		\caption{Hop distance.}
		\label{fig:hopDisN}
\end{figure}

Figure~\ref{fig:hopDisN} compares the three approaches based on the hop distance from the user to the placed services.

\paragraph{SMALL} Figure~\ref{fig:hopDisN:b} shows that the multilayer and resource-aware approaches  placed $43$, respectively $31$ services at the zero hop distance, while the availability-aware approach did not place any services at the zero hop distance. Additionally, the multilayer method placed only one service at the maximum hop distance of $257$, while the availability-aware approach placed five services at a maximum hop distance of $257$. In contrast, the resource-aware approach performed worst and placed two services at a maximum hop distance of $313$.

\paragraph{MEDIUM} Figure~\ref{fig:hopDisN:d} shows that the multilayer method placed $155$ services at the zero hop distance. In comparison, both resource and availability-aware approaches placed only $35$, respectively $1$ services at zero hop distance. Moreover, the multilayer method only places one service at the maximum hop distance of $264$, the availability-aware and resource-aware placed $8$, and $9$ services at a maximum hop distance of $244$, respectively.
\paragraph{LARGE} Figure~\ref{fig:hopDisN:e} shows that the multilayer method performed again the best and placed $68$ services at zero hop distance and used a maximum hop distance of $250$. In contrast, the availability-aware and resource-aware approaches placed $54$ and $2$ services at the zero hop distance, and used a maximum hop distance of $258$ and $303$, respectively.
\paragraph{Summary} We conclude that the multilayer method places a larger number of services across the Fog devices in close proximity to users compared to the related methods. The advantage comes from considering the transmission time between users and the Fog devices and by placing the requested services across highly connected devices.

\subsection{Deadline satisfaction}

\subsubsection{Reliable Fog infrastructure}
\begin{figure}[t]
\centering
   \begin{subfigure}[t]{.49\columnwidth} 
   \centering
       \includegraphics[width=\linewidth]{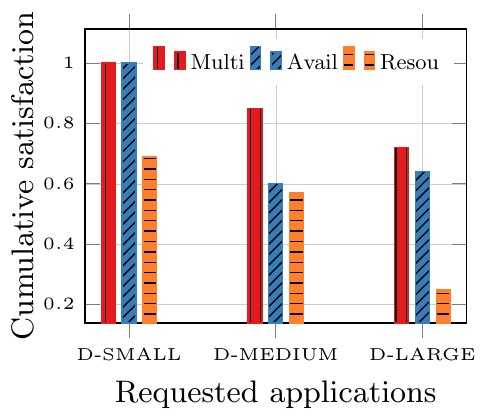}
       \vspace*{-0.6cm}
		   \caption{Reliable Fog infrastructure.}
		   \label{fig:DeadNF}
		\end{subfigure}
       \begin{subfigure}[t]{.49\columnwidth} 
       \centering
       \includegraphics[width=\linewidth]{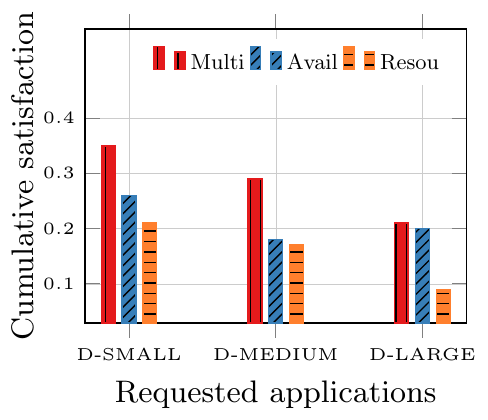}
       \vspace*{-0.6cm}
		   \caption{Faulty Fog infrastructure.}
		   \label{fig:DeadF}
		\end{subfigure}
		\caption{Deadline satisfaction.}
		\label{fig:Dead}
\end{figure}

Figure~\ref{fig:DeadNF} shows the cumulative deadline satisfaction ratio in the three \SI{2000}{\second} long simulation scenarios, specified in Table~\ref{tab:eval-scenarios}. Each user requests a random applications every \SI{1.557}{\second} (see Table~\ref{tab:cliSetup}) until the complete simulation period.

\paragraph{D-SMALL} The multilayer and availability-aware methods fulfilled the deadlines of all applications, while the resource-aware approach had a lower cumulative deadline satisfaction rate of $0.69$.

\paragraph{D-MEDIUM} The availability-aware and resource-aware approaches fulfilled deadlines with a cumulative satisfaction rate of $0.60$ and $0.57$, respectively. The multilayer method outperformed both approaches with a high cumulative deadline satisfaction rate of $0.85$.

\paragraph{D-LARGE}
The multilayer method satisfied deadlines with a high rate of $0.72$. The availability-aware approach exhibited a lower rate of $0.64$, while the resource-aware approach performed worst with a rate of $0.25$ only.

\paragraph{Summary}
We conclude with two observations.
\begin{enumerate}[label=(\roman*),font=\emph,wide]
    \item The multilayer and availability-aware methods fulfill deadlines with high cumulative satisfaction rate compared to the resource-aware approach that does not consider application deadline as a placement constraint. In contrast, both multilayer and availability-aware methods prioritize applications based on their deadline for optimized placement. 
    \item The multilayer method has a better deadline satisfaction rate with increasing application requests compared to the availability-aware approach due to its ability to place dependent services of the same application across highly connected Fog devices in the same network partitions. This optimizes the latency between dependent services and fulfills deadline for more applications with a better satisfaction rate.
\end{enumerate}

\subsubsection{Faulty Fog infrastructure}
Figure~\ref{fig:DeadF} evaluates the cumulative deadline satisfaction rate in the three scenarios for faulty Fog infrastructures. Similar to the evaluation of reliable Fog infrastructures, we show the cumulative satisfaction rate in the three \SI{2000}{\second} long simulation scenarios, where users requested random applications every \SI{1.557}{\second}. We introduced faults by randomly failing a Fog device every \SI{20}{\second}, such that all devices are not reachable at the end of simulation period of \SI{2000}{\second}.

\paragraph{D-SMALL} The availability-aware and resource-aware approaches fulfilled deadlines with a cumulative satisfaction rate of $0.29$, respectively $0.21$, in the presence of randomly failing Fog devices. The multilayer method performed slightly better with a cumulative satisfaction rate of $0.35$.

\paragraph{D-MEDIUM} All three approaches fulfilled the deadlines with a lower cumulative satisfaction rate than D-SMALL. However, the multilayer method performed better with a deadline cumulative satisfaction rate of $0.26$. In contrast, both availability-aware and resource-aware approaches exhibited a low deadline satisfaction rate of $0.18$ and $0.17$, respectively.

\paragraph{D-LARGE} The resource-aware approach performed worst and fulfilled deadlines with a very low cumulative satisfaction rate of $0.09$, while the multilayer and application-aware methods fulfilled deadlines with a better cumulative satisfaction ratio of $0.23$ and $0.20$, respectively.

\paragraph{Summary}
We draw two observations in Figure~\ref{fig:DeadF}.
\begin{enumerate}[label=(\roman*),font=\emph,wide]
    \item The resource-aware placement does not consider application deadline or device failures and exhibits poor cumulative deadline satisfaction rate compared to the multilayer and availability-aware methods across faulty Fog infrastructures. 
    \item The multilayer placement achieves a slightly higher cumulative deadline satisfaction rate compared to the availability-aware approach, which places dependent services on the same Fog devices and the others across weakly connected devices prone to failures. In contrast, the multilayer method places all application services across highly connected Fog devices in the same network layer partition and therefore, exhibits better deadline satisfaction rate even upon faults. 
\end{enumerate}

\section{Conclusions}
\label{sec:conclusion}
We introduced a new resource-aware method for Fog application placement, which represents heterogeneous Fog devices as a multilayer network graph and partitions them with respect to the network topology and resource characteristics. The new multilayer method places a multi-service application structured as a directed graph in two steps. The first step matches the requested applications with feature partitions based on their requirements, which overlap in the same topological partition. The second step places the services to Fog devices in the elected partitions closest to the end-users. We evaluated the multilayer resource-aware placement against two availability-aware and resource-aware state-of-the-art approaches. The results indicate that our method is able to place twice as many services, satisfy deadlines for three times as many application requests, and reduce the resource wastage by $15-32$ times compared to the related methods.

We plan to extend our multilayer resource-aware placement method to dynamically consider changing application and Fog infrastructure characteristics.

\section*{Acknowledgements}
\paragraph{European Union’s Horizon 2020} research and innovation programme funded this work though two grants:
\begin{itemize}[align=left, leftmargin=*]
\item ARTICONF: ``Smart Social Media Ecosytstem in a Blockchain
Federated Environment'' (grant 825134);
\item DataCloud: ``Enabling the Big Data Pipeline Lifecycle on the Computing Continuum'' (grant 101016835).
\end{itemize}
\paragraph{Carinthian Agency for Investment Promotion and Public Shareholding (BABEG)} supported the use case application and Fog infrastructure simulation.

\bibliographystyle{IEEEtran}
\bibliography{ref}

\end{document}